\documentclass[12pt,fleqn]{article}
\usepackage[cp1251]{inputenc}
\usepackage{amsmath,amssymb,amsthm,cite}
{\theoremstyle{definition}
\newtheorem{definition}{Definition}
\newtheorem{example}{Example}
\newtheorem{remark}{Remark}

\newtheorem*{note*}{Note}
}

\newcommand{\R}{{\mathbb R}}
\newcommand{\Z}{{\mathbb Z}}
\newcommand{\C}{{\mathbb C}}

\newcommand{\pr}{\mbox{pr}}
\newcommand{\p}{\partial}

\newcommand{\vspacebefore}{\raisebox{0ex}[2.5ex][0ex]{\null}}
\newcommand{\vspacebeforeM}{\raisebox{0ex}[2.9ex][0ex]{\null}}

\newcommand{\myhline}{\\[0.8ex]\hline\vspacebefore}
\begin{document}

\begin{center}
{\Large\bf Transformation Groups on Real Plane \\[2mm]
and their Differential Invariants}
\end{center}

\begin{center}
Maryna Nesterenko
\end{center}
Institute of Mathematics, 3 Tereshchenkivs'ka Str., 01601 Kyiv-4, Ukraine\\
E-mail: maryna@imath.kiev.ua\\
2000 Mathematics Subject Classification: 34C14; 34C20; 17B66.

\vspace{-1ex}

\begin{abstract}
\noindent
Complete sets of bases of differential invariants, operators of invariant differentiation
and Lie determinants of continuous transformation groups acting on the real plane are constructed.
As a necessary preliminary, realizations of finite-dimensional Lie
algebras on the real plane are revisited.
\end{abstract}

\vspace{-1ex}

\noindent
Keywords: differential invariant,
operator of invariant differentiation, Lie determinant, Lie algebra realization.

\vspace{-1ex}

\section{Introduction}
Differential invariants emerged as one of the most important tools
in investigation of dif\-fe\-ren\-tial equations in the works of
S.~Lie. In 1884~\cite{Lie1} he proved that any
non-singular invariant system of differential equations can be
expressed in terms of differential invariants of the corresponding
symmetry group. In the same paper he also applied differential
invariants to integration of ODEs.
If differential invariants of a Lie group are known,
the differential equations admitting this group can be easily described
and the special representation (so-called group
foliation) of such differential equations can be constructed.

Differential invariants of all finite-dimensional local
transformation groups on a space of two complex variables were
described by S.~Lie himself in~\cite{lie1883}. A modern
treatment of these results was adduced in~\cite{Olver1}.
Namely, functional bases of differential invariants, operators of invariant differentiation and
Lie determinants were constructed for all inequivalent realizations of
point and contact finite-dimensional transformation groups on the complex plane.
The real finite-dimensional Lie algebras of contact vector fields and their differential invariants
were completely classified in~\cite{Dubrov}.
Differential invariants of an one-parameter group of local transformations in the case
of arbitrary number of dependent and independent variables were studied in~\cite{Popovych&Boyko}.

The subject of this paper is exhaustive description of differential invariants
and Lie determinants of finite-dimensional Lie groups acting on the real plane.
A necessary prerequisite to do it is classification of Lie algebra
realizations in vector fields on the real plane up to local diffeomorphisms.

Realizations of Lie algebras by vector fields are widely applicable in the general theory of differential equations,
integration of differential equations and their systems~\cite{Olver0, Ovsyannikov},
in group classification of ODEs and PDEs~\cite{basarab-horwath&lahno&zhdanov2001},
in classification of gravity fields of a general form with respect to motion groups~\cite{petrov1966},
in geometric control theory and in the theory of systems
with superposition principles~\cite{Carinena,Shnider_Winternitz1984}.
Such realizations are also applicable in the difference schemes for
numerical solutions of differential equations~\cite{BourliouxCyr-GagnonWinternitz}.
Description of realizations is the first step for solving the Levine's problem~\cite{Levine}
on the second order time-independent Hamiltonian operators which lie in the universal enveloping algebra of
a finite-dimensional Lie algebra of the first-order differential operators.
The Levine's problem was posed in molecular dynamics.
In such a way, realizations are relevant in the theory of quasi-exactly solvable problems of quantum mechanics through
the so-called algebraic approach to scattering theory and molecular dynamics.
The list of possible applications of realizations of Lie algebras
is not exhausted by the above-mentioned subjects.

The plan of the paper is following.
In Section 2 we discuss and compare different classifications of realizations of
finite-dimensional Lie algebras on the real and complex planes, which are available in literature.
In particular, we thoroughly study the question of parametrization and equivalence in series of realizations.
The realizations of finite-dimensional Lie algebras in vector fields on the real plane
are arranged in the form of Table 1.
The transformations that reduce real Lie algebras to complex ones are presented in Table 2.
In Section 3 some definitions and results concer\-ning differential invariants are collected and
detailed example of calculation is adduced.
Using the results of Table 1 and technique proposed in Section 2, we obtain complete sets of bases
of differential invariants, operators of invariant differentiation and Lie determinants
and collect them in Table 3.
Short overview of the obtained results as well as their possible applications and development
are presented in the conclusion.

\section{Realizations of Lie algebras \\ on real and complex planes}

There are two important classification problems among a variety of others in the classical theory
of Lie algebras.

The~first one is classification of Lie algebra structures,
i.e.\ classification of possible commutation relations between basis elements.
A list of isomorphism classes of the Lie algebras is in use of many authors for different purposes e.g.
\cite{Andrada&Barberis&Dotti&Ovando2005, basarab-horwath&lahno&zhdanov2001, Burde&Steinhoff1999,
Fialowski2005, PateraSharpWinternitz1976, Patera&Winternitz1977, petrov1966}.
But the problem of unification and correction of the existing lists
(see for example~\cite{Bianchi, deGraaf, lie1888_1890_1893, MacCallum, morozov1958, mubarakzyanov1963.1,
mubarakzyanov1963.2, mubarakzyanov1963.3, Patera1990, turkowski1990})
is a very laborious task, even in the case of low dimensions,
because the number of entries in such lists rapidly increases with growing dimension
and the problem of classification of Lie algebras includes a subproblem of reduction of pair of matrices
to a canonical form \cite{Kirilov}.
Here we only remind that all possible complex Lie algebras of dimensions no greater than four were listed by S.~Lie
himself~\cite{lie1888_1890_1893} and later the semisimple Lie algebras~\cite{Jacobson}
and the Lie algebras of dimensions no greater than six \cite{morozov1958, mubarakzyanov1963.1, mubarakzyanov1963.2,
mubarakzyanov1963.3, turkowski1990} over the complex and real fields were classified.

S.~Lie established the other problem of description of different Lie algebra representations
and realizations, particularly, by vector fields up to local diffeomorphisms.

Realizations of Lie algebras by vector fields in one real, one and two complex
variables were classified by S.~Lie~\cite{lie1888_1890_1893}.
In 1990 A.~Gonzalez-Lopez et al. ordered the Lie's classification of realizations of complex Lie algebras~\cite{Olver4}
and extended it to the the real case~\cite{Olver3}.
A complete set of inequivalent realizations of real Lie algebras of dimension no greater than four
in vector fields on a space of an arbitrary (finite) number of variables was constructed in~\cite{Popovych}.
The mentioned works do not exhaust all papers devoted to realizations of Lie algebras,
but only them will be used in the present paper.

An extended overview on both these subjects is contained in the preprint math-ph/0301029v7.

Realizations of Lie algebras by vector fields in one real, one and two complex
variables were classified by S.~Lie~\cite{lie1888_1890_1893}.
In 1990 A.~Gonzalez-Lopez et al. ordered the Lie's classification of realizations of complex Lie algebras~\cite{Olver4}
and extended it to the the real case~\cite{Olver3}.
Starting from the above results,
we detailed and amended the classification of realizations of finite-dimensional Lie algebras
on the real plane.
The obtained classification is compared in Table~1 with existing classifications on real~\cite{Olver3}
and complex~\cite{lie1888_1890_1893} planes.

The nontrivial transformations over the complex field that reduce
realizations  from \cite{Olver3} to realizations from \cite{Olver1} are adduced in Table~2.

\medskip

\noindent {\bf Notations.}
Below we denote $\p/\p_x$, $\p/\p_y$, \ldots\ as $\p_x$, $\p_y$, \ldots\,.
The indices $i$ and $j$ run from 1 to~$r$, where variation range for $r$
is to be determined additionally in each case.
The label $N_0$ consists of two parts which denote
the page (from 57 to 73) and realization numbers in~\cite{lie1888_1890_1893} correspondingly.
The labels $N_1$ and $N_2$ coincide with the numerations of real and complex realizations in~\cite{Olver3,Olver1}.
$N_3$ corresponds to the numeration of realizations introduced in~\cite{Popovych}, namely $R(A,n)$
denotes the $n$-th realization of the Lie algebra $A$ from~\cite{Popovych},
or, if the dimension of the algebra is larger then four, the corresponding dimension is indicated in the
column entitled $N_3$.
The symbol~$N$ without subscripts correspond to the numeration used in the present paper.

\begin{remark}
The realization  of rank two of the non-Abelian two-dimensional real Lie algebra
$\langle \p_x, x\p_x+y\p_y\rangle$ (case $N=4$) 
is missed in~\cite{Olver3} from the formal point of view.
But it can be joined to the realization series
$\langle \p_x, \p_y, x\p_x+cy\p_y, x\p_y, \dots, x^r\p_y\rangle$, $r\ge 1$ (case $N_1=24$)
written in the form $\langle \p_x, x\p_x+cy\p_y, x^k\p_y, k=-1,0,\ldots,r\rangle$
under the supposition that for $k=-1\colon$ $x^{-1}=0$ and $c=1$.
\end{remark}

\begin{remark}
There are two different approaches to classification of
Lie algebra realizations by vector fields.
According to the first approach, one should start from classification
of Lie algebras and then look for basis vector fields
that satisfy the given commutation relations.
The second approach consists in direct construction of finite-dimensional spaces of vector fields,
which are closed with respect to the standard Lie bracket.
If a complete list of realizations of a fixed dimension is constructed
then the problem of separation of the realizations for a given Lie algebra
from others arises and becomes nontrivial in the case of parameterized series of realizations.
\end{remark}

\begin{example}
Consider the series~$\{A_{4.8}^b\}$~\cite{mubarakzyanov1963.1}
of real four-dimensional Lie algebras parameterized with the parameter $|b|\le1$.
For a fixed value of~$b$, the basis elements of~$A_{4.8}^b$ satisfy the canonical commutation relations
\[
[e_2,e_3]=e_1,\quad [e_1,e_4]=(1+b)e_1,\quad [e_2,e_4]=e_2,\quad [e_3,e_4]=be_3.
\]
In the framework of the first approach we obtain
two inequivalent realizations in vector fields on a space of two variables~\cite{Popovych}
\begin{equation}\label{eq1}
\langle \p_x,\ \p_y,\ y\p_x,\ (1+b)x\p_x+y\p_y\rangle
\quad\mbox{and}\quad
\langle \p_x,\ y\p_x,\ -\p_y,\ (1+b)x\p_x+by\p_y\rangle
\end{equation}
of the algebra~$A_{4.8}^b$ if $|b|<1$.
There is a unique inequivalent realization in the case $b=\pm1$ since under this condition
realizations~\eqref{eq1} are equivalent and we have to choose only one of them.

S.~Lie~\cite{lie1888_1890_1893} used the second approach to construct
all possible realization of finite-dimensional Lie algebras on the plane.
The algebras from the series~$\{A_{4.8}^b\}$ are represented in the obtained list by the following realizations
\begin{equation}\label{eq3}
\langle \p_y, \p_x,  x\p_y, x\p_x+\tilde b y\p_y \rangle,\ \tilde b\in \R,
\quad\mbox{and}\quad
\langle \p_y, -x\p_y, \p_x,  y\p_y\rangle.
\end{equation}
In fact, the sets of realizations~\eqref{eq1} and~\eqref{eq3} coincide.
To show it, we redenote the variables~$x$ and~$y$ in~\eqref{eq3} (namely, $x\leftrightarrow y$) at first
and  then shift the parameter~$\tilde b$: $\tilde b=1+b'$.
After reordering the basis in the first realization from~\eqref{eq3} in the case $|b'|\le1$,
we obtain the first realization from~\eqref{eq1}, where $b=b'$.
If $|b'|>1$, the first realization from~\eqref{eq3} is reduced to the second realization from~\eqref{eq1}
with $b=1/b'$ by the additional simultaneous transformations of the basis and realization variables:
$\tilde e_1=b'e_1$, $\tilde e_2=e_3$, $\tilde e_3=-b'e_2$, $\tilde e_4=be_4$;
$\tilde x=bx$, $\tilde y=by$.
The second realization from~\eqref{eq3} coincides with the second one from~\eqref{eq1}, where~$b=0$.

The above consideration explains in some way why the parameter values~$b=\pm1$ are singular for
the Lie algebra series~$\{A_{4.8}^b\}$ from the viewpoint of number of realizations.
\end{example}

\begin{remark}
It is clear that realizations from different series adduced in Table 1 are inequivalent each to other
but there can exist equivalent realizations belonging to the same series.
\end{remark}

\begin{example}
Consider the series of realizations
\begin{gather}
\label{rl1}
N=6,23,48:\quad
\langle \xi_1(x)\p_y,\ \xi_2(x)\p_y,\ \dots , \ \xi_{r+2}(x)\p_y\rangle,
\quad r\ge 1,
\\[0.5ex] \label{rl2}
N=35,49:\quad
\langle y\p_y,\ \xi_1(x)\p_y,\ \xi_2(x)\p_y,\ \dots , \ \xi_{r+2}(x)\p_y\rangle,
\quad r\ge 1
\end{gather}
parameterized with arbitrary linearly independent real-valued functions~$\xi_i$.

Any realization from series \eqref{rl1} or \eqref{rl2} pass into realizations from the same series
under the basis transformations with non-singular constant matrices $(c_{ij})$
and the non-singular variable transformations $\tilde x=\varphi(x)$, $\tilde y=\psi(x)y$.
By means of these equivalence transformations the parameter-functions~$\xi_i$ change in the following way
$\tilde \xi_i(\tilde x)=c_{ij}\psi(x)\xi_j(x)|_{\,\tilde x=\varphi(x)}$.
Consequently, without loss of generality we can put $\tilde \xi_{r+1}=1$ and $\tilde \xi_{r+2}=\tilde x$.
Hence, the series of realizations \eqref{rl1} and \eqref{rl2}
takes the form adduced in the Table 1, namely
\begin{equation}
\label{rl11}
\langle \p_{\tilde y},\ \tilde x\p_{\tilde y},\
\tilde \xi_1(\tilde x)\p_{\tilde y},\ \dots , \ \tilde \xi_r(\tilde x)\p_{\tilde y}\rangle
\quad {\rm and} \quad
\langle \tilde y\p_{\tilde y},\ \p_{\tilde y},\ \tilde x\p_{\tilde y},\
\tilde \xi_1(\tilde x)\p_{\tilde y},\ \dots , \ \tilde \xi_r(\tilde x)\p_{\tilde y}\rangle.
\end{equation}
Accurately speaking, the series with normalized forms \eqref{rl11} also contain equivalent realizations,
and the corresponding equivalence transformations are restrictions of the aforesaid ones.
\end{example}

\begin{example}
Another example is given by two series of realizations
\begin{gather}\label{rl3}
N=50:\quad
\langle \p_x, \eta_1(x)\p_y,\  \dots,\  \eta_r(x)\p_y\rangle,
\quad r\ge 4,
\\[1ex]\label{rl4}
N=51:\quad
\langle \p_x, y\p_y, \eta_1(x)\p_y,\  \dots,\  \eta_r(x)\p_y\rangle,
\quad r\ge 3
\end{gather}
parameterized with real functions $\eta_i$ which
form a fundamental system of solutions for
an $r$-order ordinary differential equation with constant coefficients
\[
\eta^{(r)}(x)+c_1\eta^{(r-1)}(x)+ \dots +c_r\eta(x)=0.
\]
The transformations that reduce any realization from the series $N=50$ and $N=51$
to a~realization from the same series are generated by
the changes of basis with non-singular constant matrices $(c_{ij})$
and the variable transformations
$\tilde x=a_1x+a_0$, $\tilde y=by+f(x)$,
where $f(x)=b_0\eta_0(x)+b_1\eta_1(x)+ \dots +b_r\eta_r(x)$,
$a_1, a_0, b, b_0, \dots, b_r \in \R$, $a_1b\ne0$.
The function~$\eta_0(x)$ is a solution of the ODE
$\eta_0^{(r)}(x)+c_1\eta_0^{(r-1)}(x)+ \dots +c_r\eta_0(x)=1$
and in the case $N=51$ additionally $b_0=0$.
These equivalence transformations act on the functions $\eta_i$ as follows:
$\tilde \eta_i(a_1x+a_0)=c_{ij}\eta_j(x)$.
\end{example}

\section{Differential invariants}

Foundations of the theory of differential invariants were established in classical works of Lie, Tresse~\cite{tresse1894, tresse1896} and Cartan
and are developed in our days~\cite{Fels1,Olver1,Olver2,Ovsyannikov}.
We shortly formulate necessary definitions and statements following~\cite{Olver2} and \cite{Ovsyannikov}
in general outlines.

Consider a local $r$-parametric transformation group $G$ acting on $M\subset X\times Y=\R\times\R$
and denote a prolonged transformation group acting on the subset of jet space
$M^{(n)}=M\times\R^n$ as $\mbox{pr}^{(n)}G$ .
Let ${\mathfrak g}$ be the $r$-dimensional Lie algebra with basis of infinitesimal operators
$\{e_i=\xi_i(x,y)\p_x+\eta_i(x,y)\p_y\}$ which corresponds to~$G$.
Then the prolonged algebra $\pr^{(n)}{\mathfrak g}$ is generated by the prolonged
first-order differential operators~\cite{Olver2, Ovsyannikov}:
\[
e_i^{(n)}=\xi_i(x,y)\p_x+\eta_i(x,y)\p_y+\sum_{k=1}^{n}\eta_i^{k}(x,y_{(k)})\p_{y^{(k)}}.
\]
Hereafter $n,k \in \mathbb{N}$, $i=1, \dots, r$,
the symbol $y_{(k)}$ denotes the tuple $(y, y', \dots, y^{(k)})$
of the dependent variable~$y$
and its derivatives with respect to~$x$ of order no greater than~$k$.

\begin{definition}
A smooth function $I=I(x,y_{(n)})\colon M^{(n)}\rightarrow \R$ is called a \emph{differential invariant}
of order~$n$ of the group $G$ if $I$ is an invariant of the prolonged group $\mbox{pr}^{(n)}G$, namely
\[
I(\mbox{pr}^{(n)}g\cdot(x,y_{(n)}))=I(x,y_{(n)}),\quad
(x,y_{(n)})\in M^{(n)}
\]
for all $g \in G$ for which $\mbox{pr}^{(n)}g\cdot(x,y_{(n)})$ is
defined.
\end{definition}

In infinitesimal terms, $I(x,y_{(n)})$ is an $n$-th order dif\-fe\-ren\-tial invariant
of the group $G$ if $e_i^{(n)}I(x,y_{(n)})=0$ for any prolonged basis infinitesimal generators~$e_i^{(n)}$ of
$\mbox{pr}^{(n)}{\mathfrak g}$.

Consider the series of the ranks
$r_k=\mbox{rank}\{(\xi_i, \eta_i, \eta_i^1, \dots, \eta_i^{k}),\ i=1, \dots, r\}$.
For further statements we introduce the number $\nu=\min\{k\in \Z\, |\, r_k=r\}$.
Since the sequence $\{r_k\}$ is nondecreasing, bounded by $r$ and reaches the value $r$,
the number $\nu$ exists and the relation $r_\nu=r_{\nu+1}=\dots=r$ holds true.

\begin{definition}
Let $\pr^{(\nu)}{\mathfrak g}$ is generated by the set of the prolonged infinitesimal operators\
$\{e_i^{(\nu)}\}$ and $L$ is the matrix formed by their coefficients:
\[
L=\left(\begin{array}{ccccc}
\xi_1 &\eta_1&\eta_1^1&\ldots&\eta_1^{(\nu)} \\
\xi_2 &\eta_2&\eta_2^1&\dots &\eta_2^{(\nu)}\\
\vdots&\vdots&\vdots  &\ddots&\vdots\\
\xi_r &\eta_r&\eta_r^1&\ldots&\eta_r^{(\nu)} \\
\end{array}\right).
\]
A maximal minor of $L$, which does not vanish identically, is called a \emph{Lie determinant}.
\end{definition}

Importance of the adduced notions is explained by the following fact.
If a system of ordinary differential equations is invariant under action of
the prolonged group $\mbox{pr}^{(n)}G$ then it can be
locally presented as a union of conditions of vanishing Lie determinants
and equations written in terms of differential invariants of $G$.

A natural question is whether it is possible to choose a minimal set of
differential invariants that allows us to obtain all differential invariants of the given order
by a finite number of certain operations. The answer to this question is affirmative.

Below we will briefly state several results concerning differential invariants of Lie groups acting on the plane.
The presented statements are well known and adduced only on purpose to complete the picture
of differential invariants on the plane.
Detailed definitions and statements on the theory of differential invariants in general cases,
review of main results
and approaches to differential invariants that are different
from the presented one
(such as differential one-forms and moving coframes)
and possibly more convenient for other applications
could be found e.g. in~\cite{Fels1, Olver2, Ovsyannikov}.

\begin{definition}
A maximal set $I_n$ of functionally independent differential invariants of order no greater than $n$
(i.e.~invariants of the prolonged group $\mbox{pr}^{(n)}G$) is called
a~\emph{universal differential invariant} of order~$n$ of the group $G$.
\end{definition}

Note that dimension of the jet space~$M^{(n)}$ is $\dim M^{(n)}=n+2$
and the number of functionally independent differential invariants
of order~$n$ is $d_n=n+2-r_n$.
Any $n$-th order differential invariant~$I$ of~$G$
is necessarily an $(n+l)$-th order differential invariant of~$G$, $l\ge 0$.
Therefore for any $n,l\ge 0$ a universal differential invariant~$I_{n+l}$ can be obtained by extension of
a universal differential invariant~$I_n$.
%\end{proposition}

\begin{definition}
A vector field (or a differential operator) $\delta$ on the infinitely prolonged jet space $M^{(\infty)}$ is
called an \emph{operator of invariant differentiation} of the group $G$
if for any differential invariant~$I$ of~$G$ the expression $\delta I$ is also a differential invariant of $G$.
\end{definition}

Any operator $\delta$ commuting with all formally infinitely prolonged basis infinitesimal
generators $e_i^{\infty}$ of the corresponding Lie algebra is an operator of invariant differentiation of $G$.
For any Lie group acting on the real or complex planes, there exists exactly one
independent (over the field of invariants of this group) operator of invariant differentiation.

For any Lie group $G$ there exists a finite \emph{basis of differential invariants},
i.e.\ a finite set of functionally independent differential invariants
such that any differential invariant of~$G$
can be obtained from it via a finite number of functional operations and operations of
invariant differentiation. A basis of differential invariants of the group~$G$ is always contained
in a universal differential invariant $I_{\nu+1}$.

To describe completely differential invariants of all transformation groups acting on the real plane,
we obtain a functional basis of differential invariants and operators of invariant
differentiation for each algebra from the known list of inequivalent Lie algebras of vector fields on the plane.

Bases of differential invariants are constructed as a part of $(\nu+1)$-th order universal differential invariants
using the infinitesimal approach.
The constructive procedure for finding invariant differentiation operators is directly derived from
the condition of their commutation with formally infinitely prolonged elements of the algebra.
Namely, we look for an operator of invariant differentiation as the operator of total differentiation~$D_x$
with a multiplier~$\lambda$ depending on~$x$ and~$y_{(\nu)}$:
$X=\lambda(x, y_{(\nu)})D_x,$
where
$\lambda\colon M^{(\nu)}\rightarrow \R.$
The function $\lambda$ is implicitly determined by the equation $\varphi(x, y_{(\nu)}, \lambda)=0$,
where~$\varphi$ satisfies the condition:
\[
\overline{\zeta_i^\nu}\varphi=0,
\qquad
\overline{\zeta_i^\nu}=\xi_i\p_x+\eta_i\p_y+\eta_i^1\p_{y'}+ \cdots+\eta_i^{\nu}\p_{y^{(\nu)}}
+(\lambda D_x)\xi_i\p_{\lambda}.
\]
In other words, $\varphi(x, y_{(\nu)}, \lambda)$ should be an invariant of the flows generated by vector fields $\overline{\zeta_i^\nu}$.
Let us note that $\mbox{rank}\{(\overline{\zeta_i^\nu}), i=1, \dots, r\}=r$.
A~universal invariant~$\overline I$ of~$\overline{\zeta_i^\nu}$ can be presented as
$\overline I= (I_\nu,\hat I)$, where $\hat I\colon M^{(\nu)}\times\R\rightarrow\R$, $\p\hat I/\p\lambda\not=0$.
So, the unknown function $\lambda$ can be find from the condition
$\hat I(x, y_{(\nu)}, \lambda)=C$ for a fixed constant~$C$.

%\pagebreak

All the obtained results obtained are presented in the form of Table 3
and may be used for group classification of ODEs of any finite order.
In a similar way one can
describe the differential invariants of the transformations groups acting in the spaces
of more than two variables and having a low number of parameters
by means of using the classification of realizations of real low-dimensional
Lie algebras~\cite{Popovych} and then apply them to investigation of systems of ODEs or PDEs.

\begin{example}
Let us illustrate the above statements by considering the Lie algebra $A_{3.5}^b$, $b\ge0$,
generated by the basis elements (case $N=17$ in Tables 1 and 3):
\begin{gather*}
e_1=\p_y,\quad e_2=x\p_y,\quad e_3=-(1+x^2)\p_x+(b-x)y\p_y.
\end{gather*}

The second prolongations of these operators are:
\begin{gather*}
e_1^{(2)}=\p_y,\\
e_2^{(2)}=x\p_y+\p_{y'},\\
e_3^{(2)}=-(1+x^2)\p_x+(b-x)y\p_y-(y-(b+x)y')\p_{y'}+(b+3x)y''\p_{y''}.
\end{gather*}

Inasmuch as dimension of the algebra is $r=3$ and the rank of the first prolongation is $r_1=3$,
the ranks of all other prolongations also equal to $3$:  \mbox{$r_1=r_2=\dots =r=3$}.
In this case $\nu=1$. Hence, the basis of differential invariants is contained in the universal differential
invariant $I_{\nu+1}=I_2$. The Lie determinant $\mathcal{L}$ is calculated as
the determinant of the matrix formed by coefficients of~$e_i^{(2)}$, $i=1,2,3$:
\begin{gather*}
\mathcal{L}=\det\left(
\begin{array}{ccc}
0 & 1 & 0\\
0 & x & 1\\
-(1+x^2) & (b-x)y & -(y-(b+x)y')
\end{array}
\right)=-(1+x^2).
\end{gather*}
It produces no invariant differential equations.

Let us look for a basis of differential invariants.
There are no differential invariants
of orders $0$ and $1$ (because of $d_0=0+2-2=0$ and $d_1=1+2-r_1=0$).
The universal differential invariant $I_2$, as well as the basis of differential invariants,
is formed by a single (because of $d_2=2+2-r_2=1$) function $I=I(x,y,y',y'')$.
It is defined by the conditions $e_i^{(2)}I=0,$ $i=1,2,3$,
which are equivalent to the overdetermined system of first-order linear PDEs
\begin{gather*}
\frac{\p I}{\p y}=0,\\
x\frac{\p I}{\p y}+\frac{\p I}{\p y'}=0,\\
-(1+x^2)\frac{\p I}{\p x}+(b-x)y\frac{\p I}{\p y}-(y-(b+x)y')\frac{\p I}{\p y'}+(b+3x)y''\frac{\p I}{\p y''}=0.
\end{gather*}

It is follows from two first equations that $I=I(x,y'')$.\
The basis of differential invariants of the considered realization is obtained
as the set of functionally independent integrals for the reduced characteristic system of the third equation:
\[I_2=\{y''(1+x^2)^{\frac 32}e^{b\arctan x}\}.\]

The last task which should be solved in order to describe all differential invariants
of this realization is construction of an operator of invariant differentiation.

We look for the operator of invariant differentiation in the form $\lambda(x,y,y')D_x$.
Here $D_x$ is the operator of total differentiation.
The function $\lambda$ is implicitly given by the equation $\varphi(x,y,y',\lambda)=0$,
where $\varphi$ is a nonconstant solution of the overdetermined system of first-order linear PDEs
\begin{gather*}
\frac{\p \varphi}{\p y}=0,\quad
x\frac{\p \varphi}{\p y}+\frac{\p \varphi}{\p y'}=0,\\
(1+x^2)\frac{\p \varphi}{\p x}-(b-x)y\frac{\p \varphi}{\p y}
+(y-(b+x)y')\frac{\p \varphi}{\p y'}+2x\lambda\frac{\p \varphi}{\p \lambda}=0.
\end{gather*}

Two first equations result in the condition $\varphi=\varphi(x,\lambda)$. Then the latter implies
$\varphi=\varphi(\omega)$, where $\omega=\lambda(1+x^2)^{-1}$, i.e.
the function $\lambda$ can be found from the equation \mbox{$\lambda(1+x^2)^{-1}=1$}.
The corresponding operator of invariant differentiation is $(1+x^2)D_x$.

\end{example}
\begin{remark}
Form of differential invariants essentially depends on explicit form of realizations.
Therefore, to construct an optimal set of invariants, one should to choose an optimal form of realizations.
For example, for the realization
$
\langle e^{-bx}\sin{x}\p_y,\ e^{-bx}\cos{x}\p_y,\ \p_x\rangle,
$
where $b\ge 0$ ($N=17$),
a fundamental differential invariant, an operator of invariant differentiation and a Lie determinant
have the following form:
\[
I_2=y''+2by'+(b^2+1)y,\quad X=D_x,\quad L=-e^{-2bx}.
\]
For the equivalent form
$
\langle \p_y,\  x\p_y,\ -(1+x^2)\p_x+(b-x)y\p_y \rangle
$
which is considered in Example~4,
the corresponding invariant objects are more complicated:
\begin{gather*}
I_2=y''(1+x^2)^{3/2}e^{b\arctan{x}},\quad X=(1+x^2)D_x,\quad L=-(1+x^2).
\end{gather*}
\end{remark}

\begin{remark}
The cases marked with ``$*$'' in Table 3 differ from the cases with the same numbers by changes of
dependent and independent variables.
They are adduced simultaneously because different forms may be convenient for different applications.
\end{remark}

\section{Concluding remarks}
In this paper we provide a complete description of differential invariants
and Lie determinants of finite-dimensional Lie groups acting on the real plane.
Obtained results are presented in Table 3. As preliminaries of the above problem,
known results on classification of realizations of real Lie algebras in vector fields
in two variables were reviewed and amended (Table 1). Namely, the low dimensional Lie algebras
were extracted from the general cases and presented in explicit form,
what made this classification more convenient for applications.
Additionally, the problem of equivalence of realizations belonging
to the same series of Lie algebra was discussed.

Results of the paper could be applied to group classification of ODEs
of any finite order over the real field.
In the future we plan to review and to generalize results of group classification
of the third and fourth order ODEs that were obtained in \cite{Nucci,Schmucker}.
In a similar way one can describe the differential invariants of the transformations groups
acting in the spaces of more than two variables and having
a low number of parameters by means of using
the classification of realizations of real low-dimensional
Lie algebras and then apply them to investigation of systems of ODEs or PDEs.

\newpage

\begin{center}
{\bf Table 1.} Realizations of Lie algebras on the real plane.
\vspace{0.5 cm}
\small\renewcommand{\arraystretch}{1.2}
\begin{tabular}{|c|l|c|l|l|}
%\begin{tabular}{|@{\,\,}c@{\,}|@{\,\,}l@{\,}|@{\,}c@{\,}|@{\,}l@{\,}|@{\,}l@{\,}|}
\hline\vspacebefore
$N$ &\hfil Realizations& \hfil $N_1$ &\hfil $N_0$ &\hfil $N_3$\\
\hline\vspacebefore
1&
$\p_x$ &9 & 57,\,(1)&$R(A_1,1)$
\myhline
2&
$\p_x$, $\p_y$ &22 & 57,\,(2)&$R(2A_1,1)$
\myhline
3&
$\p_x$, $y\p_x$ &20 &57,\,(4) &$R(2A_1,2)$
\myhline
4&
$\p_x$, $x\p_x+y\p_y$&--- & 57,\,(3)&$R(A_{2.1},1)$
\myhline
5&
$\p_x$, $x\p_x$&10 & 57,\,(5)&$R(A_{2.1},2)$
\myhline
6&
$\p_y$, $x\p_y$, $\xi_1(x)\p_y$ &20 & 57,\,(14)&$R(3A_1,5)$
\myhline
7&
$\p_y,y\p_y$, $\p_x$&23&73,\,(10)&$R(A_{2.1}\oplus A_1,3)$
\myhline
8&
$e^{-x}\p_y$, $\p_x$, $\p_y$&22&57,\,(8)&$R(A_{2.1}\oplus A_1,4)$
\myhline
9&
$\p_y$, $\p_x$, $x\p_y$&22&57,\,(9)&$R(A_{3.1},3)$
\myhline
10&
$\p_y$, $\p_x$,  $x\p_x+(x+y)\p_y$&25&57,\,(11)&$R(A_{3.2},2)$
\myhline
11
&$e^{-x}\p_y$, $-xe^{-x}\p_y$, $\p_x$&22&57,\,(7)&$R(A_{3.2},3)$
\myhline
12&
$\p_x$, $\p_y$, $x\p_x+y\p_y$&12&57,\,(10)&$R(A_{3.3},2)$
\myhline
13&
$\p_y$, $ x\p_y$, $y\p_y$&21&57,\,(15)&$R(A_{3.3},4)$
\myhline
14&
$\p_x$, $\p_y$, $x\p_x+ay\p_y$, $0<|a|\le 1, a\ne 1$&12&57,\,(10)&$R(A_{3.4}^a,2)$
\myhline
15&
$e^{-x}\p_y$, $e^{-ax}\p_y$, $\p_x$, $0<|a|\le 1, a\ne 1$&22&57,\,(6)&$R(A_{3.4}^a,3)$
\myhline
16&
$\p_x$, $\p_y$, $(bx+y)\p_x+(by-x)\p_y$,
$b\ge 0$&1&$\overset{\C}{\sim}57,\,(10)$&$R(A_{3.5}^b,2)$
\myhline
17&
$e^{-bx}\sin{x}\p_y,\ e^{-bx}\cos{x}\p_y,\ \p_x,\ b\ge 0$
&22&$\overset{\C}{\sim}57,\,(6)$&$R(A_{3.5}^b,3)$
\myhline
18&
$\p_x$, $x\p_x+y\p_y$, $(x^2-y^2)\p_x+2xy\p_y$ &2&
$\overset{\C}{\sim}57,\,(13);\,73,\,(4)$&$R(sl(2,{\mathbb R}),2)$
\myhline
19&
$\p_x+\p_y$, $x\p_x+y\p_y$,
$x^2\p_x+y^2\p_y$ &17&57,\,(13);\,73,\,(4)&$R(sl(2,{\mathbb R}),3)$
\myhline
20&
$\p_x$, $x\p_x+\frac 12 y\p_y$,
$x^2\p_x+xy\p_y$ &18&57,\,(16);\,72,\,(10)&$R(sl(2,{\mathbb R}),4)$
\myhline
21&
$\p_x$, $x\p_x$, $x^2\p_x$ &11&$\overset{\C}{\sim}57,\,(16);\,72,\,(10)$&$R(sl(2,{\mathbb R}),5)$
\myhline
22&
$\begin{array}{@{}l@{}}
y\p_x-x\p_y,\ (1+x^2-y^2)\p_x+2xy\p_y,\\
2xy\p_x+(1+y^2-x^2)\p_y
\end{array}$
&3&$\overset{\C}{\sim}57,\,(13);\,73,\,(4)$&$R(so(3),1)$
\myhline
23&
$\p_y,$ $x\p_y,$ $\xi_1(x)\p_y,$ $\xi_2(x)\p_y$ &20&58,\,(8)&$R(4A_1,11)$
\myhline
24&
$\p_x$, $x\p_x$, $\p_y$, $y\p_y$ &13&58,\,(6)&$R(2A_{2.1},5)$
\myhline
25&
$e^{-x}\p_y,$ $\p_x,$ $\p_y$, $y\p_y$ &23&58,\,(1)&$R(2A_{2.1},7)$
\myhline
26& $e^{-x}\p_y,$ $-xe^{-x}\p_y,$ $\p_x,$ $\p_y$ &22&57,\,(21)&$R(A_{3.2}{\oplus}A_1,9)$
\myhline
27&
$e^{-x}\p_y$, $e^{-ax}\p_y$, $\p_x$, $\p_y$, $0<|a|\le 1, a\ne 1$&22&57,\,(20)&$R(A_{3.4}^a{\oplus} A_1,9)$
\myhline
28&
$e^{-bx}\sin{x}\p_y,\ e^{-bx}\cos{x}\p_y,\ \p_x,\ \p_y,\ b\ge 0$
& 22 & $\overset{\C}{\sim}57,\,(20)$ & $R(A_{3.5}^b{\oplus}A_1,8)$
\myhline
29&
$\p_x$, $x\p_x$, $y\p_y$, $x^2\p_x+xy\p_y$ &19&58,\,(7)&$R(sl(2,{\mathbb R}){\oplus}A_1,8)\!\!$
\myhline
30&
$\p_x$, $\p_y$, $x\p_x$, $x^2\p_x$ &14&58,\,(3)&$R(sl(2,{\mathbb R}){\oplus} A_1,9)\!\!\!\!$\\
\hline
\end{tabular}
\end{center}
%\end{landscape}

\newpage
\begin{center}
{\bf Table 1.} (Continued.) \\
\vspace{0.3 cm}
\small\renewcommand{\arraystretch}{1.2}
\begin{tabular}{|c|l|c|l|l|}
\hline\vspacebefore
$N$ &\hfil Realizations&\hfil  $N_1$ &\hfil  $N_0$ &\hfil  $N_3$\\
\hline\vspacebefore
31&
$\p_y$, $-x\p_y$, $\frac 12 x^2 \p_y$, $\p_x$ &22&57,\,(23)&$R(A_{4.1},8)$
\myhline
32&
$e^{-bx}\p_y$, $e^{-x}\p_y$, $-xe^{-x}\p_y$, $\p_x$&22&57,\,(18)& $R(A_{4.2}^{b\ne1},8)$
\myhline
33&
$e^{-x}\p_y$, $-x\p_y$, $\p_y$, $\p_x$ &22&57,\,(22)&$R(A_{4.3},8)$
\myhline
34&
$e^{-x}\p_y$, $-xe^{-x}\p_y$, $\frac 12 x^2e^{-x}\p_y$,  $\p_x$ &22&57,\,(19)&$R(A_{4.4},7)$
\myhline
35&
$\p_y$, $x\p_y$, $\xi_1(x)\p_y$, $y\p_y$ &21&58,\,(9)&$R(A_{4.5}^{1,1,1},10)$
\myhline
36&
$e^{-ax}\p_y$, $e^{-bx}\p_y$, $e^{-x}\p_y$, $\p_x$, $-1\le a<b<1$, $ab\ne0$&22&57,\,(17)&$R(A_{4.5}^{a,b,1},7)$
\myhline
37&
$e^{-ax}\p_y,\ e^{-bx}\sin{x}\p_y,\ e^{-bx}\cos{x}\p_y,\ \p_x$, $a>0$
&22&$\overset{\C}{\sim}57,\,(17)$&$R(A_{4.6}^{a,b},6)$
\myhline
38&
$\p_x$, $\p_y$, $x\p_y$, $x\p_x+(2y+x^2)\p_y$ &25&58,\,(5)&$R(A_{4.7},5)$
\myhline
39&
$\p_y$, $\p_x$, $x\p_y$, $(1+b)x\p_x+y\p_y$, $|b|\le1$ &24&58,\,(4)&$R(A_{4.8}^b,5)$
\myhline
40&
$\p_y$, $-x\p_y$, $\p_x$, $y\p_y$&23&58,\,(2);\,72,\,(7)\!\!&$R(A_{4.8}^0,7)$
\myhline
41&
$\p_x$, $\p_y$, $x\p_x+y\p_y$, $y\p_x-x\p_y$ &4&$\overset{\C}{\sim}58,\,(6)$&$R(A_{4.10},6)$
\myhline
42&
$\sin{x}\p_y$, $\cos{x}\p_y$, $y\p_y$, $\p_x$&23&$\overset{\C}{\sim}58,\,(1)$&$R(A_{4.10},7 )$
\myhline
43&
$\p_x$, $\p_y$, $x\p_x-y\p_y$, $y\p_x$, $x\p_y$ & 5 & 71,\,(3)& $\dim A=5$
\myhline
44&
$\p_x$, $\p_y$, $x\p_x$, $y\p_y$, $y\p_x$, $x\p_y$ & 6 & 71,\,(2) & $\dim A=6$
\myhline
45&
$\begin{array}{@{}l@{}}
\p_x,\ \p_y,\ x\p_x+y\p_y,\ y\p_x-x\p_y,\\
(x^2-y^2)\p_x-2xy\p_y,\ 2xy\p_x-(y^2-x^2)\p_y
\end{array}$
& 7& $\overset{\C}{\sim}73,\,(3)$ & $\dim A=6$
\myhline
46&
$\p_x$, $\p_y$, $x\p_x$, $y\p_y$, $x^2\p_x$, $y^2\p_y$ &16 &73,\,(3)& $\dim A=6$
\myhline
47&
$\begin{array}{@{}l@{}}
\p_x,\ \p_y,\ x\p_x,\ y\p_y,\ y\p_x,\ x\p_y,\\
x^2\p_x+xy\p_y,\ xy\p_x+y^2\p_y
\end{array}$
& 8& 71,\,(1)&$\dim A=8$
\myhline
48&
$\p_y$, $x\p_y$, $\xi_1(x)\p_y$, $\dots$, $\xi_r(x)\p_y$, $r\ge 3$ & 20 & 73,\,(2) &$\dim A\ge5$
\myhline
49&
$y\p_y$, $\p_y$, $x\p_y$, $\xi_1(x)\p_y$, $\dots$, $\xi_r(x)\p_y$, $r\ge 2$ & 21 & 72,\,(8) & $\dim A\ge5$
\myhline
50&
$\p_x$, $\eta_1(x)\p_y$, $\dots$, $\eta_r(x)\p_y$, $r\ge4$& 22& 73,\,(1)&$\dim A\ge5$
\myhline
51&
$\p_x$, $y\p_y$, $\eta_1(x)\p_y$,  $\dots$, $\eta_r(x)\p_y$, $r\ge3$&23&72,\,(7) &$\dim A\ge5$
\myhline
52&
$\p_x$, $\p_y$, $x\p_x+cy\p_y$, $x\p_y$, $\dots$, $x^r\p_y$, $r\ge 2$&24 & 72,\,(5)&$\dim A\ge5$
\myhline
53&
$\p_x$, $\p_y$, $x\p_y$,  \dots, $x^{r-1}\p_y,x\p_x\!+\!(ry\!+\!x^r)\p_y$, $r\ge 3\!\!$ & 25 & 72,\,(6) & $\dim A\ge5$
\myhline
54&
$\p_x$,  $x\p_x$, $y\p_y$, $\p_y$, $x\p_y$, \dots, $x^r\p_y$,  $r\ge 1$&26 & 72,\,(4)&$\dim A\ge5$
\myhline
55&
$\begin{array}{@{}l@{}}
\p_x,\ \p_y,\ 2x\p_x+ry\p_y,\ x^2\p_x+rxy\p_y,\\
x\p_y,\ x^2\p_y,\ \dots,\ x^r\p_y,\ r\ge 1
\end{array}$
&27&71,\,(4);\,72,\,(1)\!\! &$\dim A\ge5$
\myhline
56&
$\begin{array}{@{}l@{}}
\p_x, x\p_x, y\p_y, x^2\p_x+rxy\p_y,\\
\p_y, x\p_y, x^2\p_y, \ \dots, x^r\p_y, r\ge 0
\end{array}$
&\!\!15;\,28\!\! &73,\,(5);\,72,\,(2)\!\! & $\dim A\ge5$\\
\hline
\end{tabular}
\end{center}
\vspace{0.5 ex}

The functions~$1$, $x$, $\xi_1$, \ldots, $\xi_r$ are linearly independent.
The functions~$\eta_1$, \ldots, $\eta_r$ form a fundamental system of solutions for
an $r$-order ordinary differential equation with constant coefficients
$\eta^{(r)}(x)+c_1\eta^{(r-1)}(x)+ \dots +c_r\eta(x)=0$.

\newpage
\begin{center}
{\bf Table 2.} Transformations of real realizations to complex ones.\\
\vspace{0.5 cm}
\small\renewcommand{\arraystretch}{1.2}
\begin{tabular}{|r|l|l|l|}
\hline
$N_1$&
\multicolumn{1}{|c|}{$\begin{array}{@{}c@{}}
\mbox{\rm Transformation}\\
\mbox{\rm of\ space\ variables}
\end{array}$}
&
\multicolumn{1}{|c|}{$\begin{array}{@{}c@{}}
\mbox{\rm Transformation\ of\ basis\ elements}
\end{array}$}
&\multicolumn{1}{|c|}{$N_2$}\\[1mm]
\hline
&&&\\[-4.5mm]
1&$\tilde x=x-iy$, $\tilde y=x+iy$&
$\tilde e_1=\frac{1+i}{2}(e_1+e_2)$, $\tilde e_2=\frac{1}{c+i}e_3$,
$\tilde e_3=\frac{1-i}{2}(e_1-e_2)$&$2.7$, $k=1$\\[1mm]
\hline
&&&\\[-4.5mm]
2&$\tilde x=x-iy$, $\tilde y=\frac{1}{2iy}$&
$\tilde e_1=e_1$, $\tilde e_2=e_2$,
$\tilde e_3=e_3$
&$2.2$\\[1mm]
\hline
&&&\\[-4.5mm]
3&$\tilde x=-\frac{1}{ix+y}$, $\tilde y=\frac{ix+y}{1+x^2+y^2}$&
$\tilde e_1=\frac{1}{2}(ie_2+e_3)$, $\tilde e_2=ie_1$,
 $\tilde e_3=\frac{1}{2}(e_3-ie_2)$&$2.2$\\[1mm]
\hline
&&&\\[-4.5mm]
4&$\tilde x=\frac{y-ix}{2}$, $\tilde y=-\frac{y+ix}{2}$&
$\tilde e_1=ie_1-e_2$,\ $\tilde e_2=ie_1+e_2$,
 $\tilde e_3=\frac{e_3+ie_4}{2}$, $\tilde e_4=\frac{e_3-ie_4}{2}$
&$2.9$, $k=1$\\[1mm]
\hline
&&&\\[-4.5mm]
7&$\tilde x=y+ix$, $\tilde y=y-ix$&
$\begin{array}{@{}l@{}}
\tilde e_1=\frac{e_1+ie_2}{2i},\ \tilde e_2=\frac{e_3-ie_4}{2},
\ \tilde e_3=\frac{e_6+ie_5}{2},\\
\tilde e_4=\frac{ie_2-e_1}{2i},\
\tilde e_5=\frac{e_3+ie_4}{2},\ \tilde e_6=\frac{e_6-ie_5}{2}
\end{array}$
&$2.4$\\[1mm]
\hline
17&$\tilde x=y$, $\tilde y=\frac{1}{x-y}$&
$\tilde e_1=e_1$, $\tilde e_2=e_2$,
$\tilde e_3=e_3$
&$2.2$\\[1mm]
\hline
&&&\\[-4.5mm]
18&$\tilde x=x$, $\tilde y=\frac{1}{y^2}$&
$\tilde e_1=e_1$, $\tilde e_2=\frac{1}{2}e_2$, $\tilde e_3=e_3$&$2.1$\\[1mm]
\hline
&&&\\[-4.5mm]
19&$\tilde x=x$, $\tilde y=\frac{1}{y}$&
$\tilde e_1=e_1$, $\tilde e_2=e_2$, $\tilde e_3=-e_3$, $\tilde e_4=e_4$&
$2.3$\\[1mm]
\hline
\end{tabular}
\end{center}
\vspace{0.2ex}

\noindent
In Table~3 we use the following notations:
\begin{gather*}
S_{k+3}=(k+1)^2\big(y^{(k)}\big)^2y^{ (k+3)}\!
-3(k+1)(k+3)y^{(k)}y^{(k+1)}y^{(k+2)}+2(k+2)(k+3)\big(y^{(k+1)}\big)^3,
\\[.8ex]
Q_{k+2}=(k+1)y^{(k)}y^{(k+2)}-(k+2)\big(y^{(k+1)}\big)^2,\qquad
\tilde{Q_3}=y'''B_1-3y'(y'')^2,
\\[.8ex]
B_0=1+x^2+y^2, \qquad B_1=1+(y')^2,
\\[.8ex]
P_{i,j}(\varphi,\psi)=\varphi^{(i)}\psi^{(j)}-\varphi^{(j)}\psi^{(i)},\qquad
R_4=3y''y^{\mbox{\scriptsize \i v}}-5(y''')^2,
\\[1ex]
\tilde{U_5}=4y^{\mbox{\scriptsize v}}B_1^3Q+10y^{\mbox{\scriptsize \i v}}y''B_1^3\left(4y'''y'+3(y'')^2\right)
-5(y^{\mbox{\scriptsize \i v}})^2B_1^4+40(y''')^2(y'')^2\left((y')^2-2\right)B_1^2
\\
\phantom{U=}{}
{}-40(y''')^3y'B_1^3-180y'''y'(y'')^4\left((y')^2-1\right)B_1^2-(y'')^6\left(45(6(y')^2+1)-135(y')^4\right),
\\[1ex]
U_5=(y')^2\left(Q_3D_x^2Q_3-\tfrac{5}{4}(D_xQ_3)^2\right)+y'y''Q_3D_xQ_3-
\left(2y'y'''-(y'')^2\right)Q_3^2,
\\[1ex]
V_7=(y'')^2\left(S_5D_x^2S_5-\tfrac{7}{6}(D_xS_5)^2\right)+y''y'''S_5D_xS_5-\tfrac{1}{2}
\left(9y''y^{\mbox{\scriptsize \i v}}-7(y''')^2\right)S_5^2,
\\[1ex]
W(f_1,f_2,\dots,f_r)=
\left|
\begin{array}{cccc}
f_1(x) & f_2(x) & \dots & f_r(x)\\
f'_1(x) & f'_2(x) & \dots & f'_r(x)\\
\dots  &\dots &\dots &\dots \\
f^{(r-1)}_1(x) & f^{(r-1)}_2(x) & \dots & f^{(r-1)}_r(x)\\
\end{array}
\right|,
\\[1ex]
K_r(\eta_1,\eta_2, \dots , \eta_r)=y^{(r)}+c_1y^{(r-1)}+c_1y^{(r-1)}+ \dots +c_ry,
\end{gather*}
where $c_1, \dots, c_r$ are the constant coefficients of the $r$-th-order ODE
\[\eta^{(r)}(x)+c_1\eta^{(r-1)}(x)+ \dots +c_r\eta(x)=0\]
which is satisfied by the functions
$\eta_1(x)$, \dots, $\eta_r(x)$.

\newpage
\begin{center}
{\bf Table 3.} Differential invariants, operators of invariant differentiation\\
and Lie determinants of realizations of Lie algebras on the real plane.\\
\vspace{0.5 cm}
\small\renewcommand{\arraystretch}{1.4}
\begin{tabular}{|l|l|l|l|}
\hline
\multicolumn{1}{|c|}{$N$}&
\multicolumn{1}{|c|}{
Basis of differential invariants}&
\multicolumn{1}{|c|}{\parbox{36mm}{\centering Operator}\rule[-2.7ex]{0pt}{6.5ex}}&
\multicolumn{1}{|c|}{Lie determinant}\\
\hline
\vspacebefore
1&
$y$&$D_x$&{\rm const}\\
\hline
1*&
$x$, $y'$&$D_x$&{\rm const}\\
\hline
2&
$y'$, $y''$ &$D_x$&{\rm const}\\
\hline
\vspacebefore
3&
$y$, $\frac{y''}{y'^3}$&$\frac{1}{y'}D_x$&$-(y')^2$\\
\hline
3*&
$x$, $y''$&$D_x$&{\rm const}\\
\hline
4&
$y'$ & $y D_x$ & $y$\\
\hline
\vspacebefore
%&&&\\[-5mm]
5&
$y$, $\frac{y''}{(y')^2}$&$\frac{1}{y'}D_x$&$y'$\\
&&&\\[-6 mm]
\hline
\vspacebefore
5*&
$x$, $\frac{y''}{y'}$&$D_x$&$y'$\\[0.4ex]
\hline
6&
$x$, $y''\xi_1'''(x)-y'''\xi_1''(x)$&$D_x$&$\xi_1''(x)$\\
\hline
\vspacebefore
7&
$\frac{y''}{y'}$&$D_x$&$y'$\\
\hline
8&
$y''+y'$&$D_x$&$-e^{-x}$\\
\hline
9&
$y''$&$D_x$&{\rm const}\\
\hline
10&
$y''e^{y'}$&$e^{y'}D_x$&{\rm const}\\
\hline
11&
$y''+2y'+y$&$D_x$&$-e^{-2x}$\\
\hline\vspacebefore
12&
$\frac{y'''}{(y'')^2}$&$\frac{1}{y''}D_x$&$-y''$\\
\hline
\vspacebefore
13&
$x$, $\frac{y'''}{y''}$&$D_x$&$-y''$\\
\hline
\vspacebefore
14&
$y''y'^{\frac{2-a}{a-1}}$ & $(y')^{\frac{1}{a-1}}D_x$&$(a-1)y'$\\
\hline
15&
$y''+(a+1)y'+ay$&$D_x$&$(1-a)e^{-(1+a)x}$\\
\hline
16&
$y''e^{-c\arctan y'}B_1^{-3/2}$&${e^{-c\arctan y'}B_1^{-1/2}}D_x$&$B_1$\\
\hline
17&
$y''+2by'+(b^2+1)y$&$D_x$&$-e^{-2bx}$\\
\hline
18&
$(y''y+(y')^2+1){B_1^{-3/2}}$&${2y}{B_1^{-1/2}}D_x$&$2y^2B_1$\\
\hline
19&
$(y''(x-y)+2y'(1+y'))(y')^{-3/2}$&$(x-y)(y')^{-1/2}D_x$&$2y'(x-y)^2$\\
\hline
20&
$y^3y''$&$y^2D_x$&$y^2$\\
\hline
\vspacebefore
21&
$x$, $(y')^{-2}{Q_3}$&$D_x$&$y(y-x)y'$\\
\hline
21*&
$y$, $(3y''^2-2y'y''')(y')^{-4}$&$\frac{1}{y'}D_x$&$y'$\\
\hline
22&
$y''B_0B_1^{-3/2}+2(y-xy')B_1^{-1/2}$&$B_0B_1^{-1/2}D_x$&$B_0^2B_1$\\
\hline
23&
$x$,
$y'''P_{2,4}(\xi_1,\xi_2)+y''P_{4,3}(\xi_1,\xi_2)+y^{\mbox{\scriptsize \i v}}$ & $D_x$ & $P_{2,3}(\xi_1,\xi_2)$\\
\hline
\vspacebefore
24&
$\frac{y'y'''}{(y'')^2}$& $\frac{y'}{y''}D_x$& $y'y''$\\
\hline
\vspacebefore
25&
$\frac{y'''+y''}{y''+y'}$&$D_x$&$-e^{-x}(y''+y')$\\
\hline
26&
$y'''+2y''+y'$&$D_x$&$-e^{-2x}$\\
\hline
27&
$y'''+(1+a)y''+ay'$&$D_x$&$a(a-1)e^{-(1+a)x}$\\
\hline
\end{tabular}
\end{center}

\newpage
\begin{center}
{\bf Table 3.} (Continued.)\\
\vspace{0.5 cm}
\small\renewcommand{\arraystretch}{1.4}
\begin{tabular}{|l|l|l|l|}
\hline
\multicolumn{1}{|c|}{$N$}&
\multicolumn{1}{|c|}{
Basis of differential invariants}&
\multicolumn{1}{|c|}{\parbox{20 mm}{\centering Operator}\rule[-2.7ex]{0pt}{6.0ex}}&
\multicolumn{1}{|c|}{Lie determinant}\\
\hline
\vspacebefore
28&
$y'''+2by''+(1+b^2)y'$&$D_x$&$-(1+b^2)e^{-2bx}$\\
\hline
\vspacebefore
29&
${S_3}{Q_2^{-3/2}}$& $\sqrt{\frac{y}{y''}}D_x$& $-2y^2y''$\\
\hline
\vspacebefore
30&
${Q_3}(y')^{-4}$& $\frac{1}{y'}D_x$& $2y'^2$\\
\hline
31&
$y'''$&$D_x$&{\rm const}\\
\hline
32&
$y'''+(b+2)y''+(2b+1)y'+by$&$D_x$&$(b-1)^2e^{-(b+2)x}$\\
\hline
33&
$y'''+y''$&$D_x$&$-e^{-x}$\\
\hline
34&
$y'''+3y''+3y'+y$&$D_x$&$-e^{-3x}$\\
\hline
35&
$x$, ${P_{2,4}(\xi_1,y)}/{P_{2,3}(\xi_1,y)}$&$D_x$&$P_{2,3}(\xi_1,y)$\\
\hline
\vspacebefore
36&
%$y'''+(a+b+1)y''+(ab+a+b)y'+aby$&$D_x$&$e^{-(a+b+1)x}$\\
$y'''+(a+b+1)y''+(ab+a+b)y'+aby$&$D_x$&$\frac{(b-a)(1-a)(1-b)}{e^{(a+b+1)x}}$\\
\hline
\vspacebefore
37&
$y'''+(2b+a)y''+(b^2+2ab+1)y'+a(b^2+1)y\!\!$&$D_x$&$((b-a)^2+1)e^{-(2b+a)x}\!\!$\\
\hline
\vspacebeforeM
38&
$y'''e^{\frac{y''}2}$&$e^{\frac{y''}2}D_x$&{\rm const}\\
\hline
\vspacebefore
39&
$b=1\colon \quad y'',\; y^{\mbox{\scriptsize \i v}}(y''')^{-2}$&$\frac1{y'''}D_x$&$y'''$\\
\cline{2-4}
%\vspacebeforeM
&
$b\ne 1\colon \quad (y'')^{\frac{2-b}{b-1}}$&$y''^{\frac1{b-1}}D_x$&$(1-b)y''$\\
\hline
\vspacebefore
40&
$\frac{y'''}{y''}$&$D_x$&$-y''$\\
\hline
\vspacebefore
41&
$(y'')^{-2}B_1y'''-3y'$& $\frac{B_1}{y''}D_x$& $3y''B_1$\\
\hline\vspacebefore
42&
$\frac{y'''+y'}{y''+y}$&$D_x$&$y''+y$\\
\hline
\vspacebeforeM
43&
$(3y''y^{\mbox{\scriptsize \i v}}-5(y''')^2)(y'')^{-8/3}$& $(y'')^{-1/3}D_x$& $y''$\\
\hline
\vspacebefore
44&
${S_5}{R_4^{-3/2}}$& ${y''}{R_4^{-1/2}}D_x$& $(y'')^2R_4$\\
\hline
\vspacebefore
45&
${\tilde U_5}\tilde Q_3^{-3}$&${B_1} \tilde Q_3^{-1/2}D_x$&$-16B_1\tilde Q_3^{2}$\\
\hline
\vspacebeforeM
46&
${U_5}{Q_3^{-3}}$&${y'}{Q_3^{-1/2}}D_x$&$-4y'Q_3^{-2}$\\
\hline
\vspacebefore
47&
${V_7}{S_5^{-8/3}}$& ${y''}{S_5^{-1/3}}D_x$& $-2y''S_5^2$\\
\hline
48&
$x$, $W(y'',\xi''_1,\xi''_2,\dots,\xi''_r)$& $D_x$& $W(\xi''_1,\xi''_2,\dots,\xi''_r)$\\
\hline
\vspacebefore
49&
$x$, $D_x\ln|{W(y'',\xi''_1,\xi''_2,\dots,\xi''_r)}|$
&$D_x$&$W(y'',\xi''_1,\dots,\xi''_r)$\\
\hline
50&
$K_r(\eta_1,\dots , \eta_r)$& $D_x$& $W(\eta_1,\eta_2,\dots,\eta_r)$\\
\hline
\vspacebefore
51&
$D_x\ln|{K_r(\eta_1,\dots , \eta_r)}|$& $D_x$& $W(y,\eta_1,\dots,\eta_r)$\\
\hline
\vspacebefore
52&
$c\ne r+1\colon \quad (y^{(r+1)})^{\frac{2-c+r}{c-r-1}}y^{(r+2)}$&$(y^{(r+1)})^{\frac{1}{c-r-1}}D_x$&$y^{(r+1)}$\\
\cline{2-4}
\vspacebefore
&
$c=r+1\colon \quad y^{(r+1)},\ \frac{y^{(r+3)}}{(y^{(r+2)})^2}$&$\frac{1}{y^{(r+2)}}D_x$&$y^{(r+2)}$\\
\hline
\vspacebeforeM
53&
$y^{(r+1)}e^{\frac{y^{(r)}}{r!}}$&$e^{\frac{y^{(r)}}{r!}}D_x$&${\rm const}$\\
\hline
\vspacebeforeM
54&
$\frac{y^{(r+1)}y^{(r+3)}}{(y^{(r+2)})^2}$& $\frac{y^{(r+1)}}{y^{(r+2)}}D_x$& $y^{(r+1)}y^{(r+2)}$\\
\hline
\vspacebefore
55&
$Q_{r+3}(y^{(r+1)})^{-\frac{2r+8}{r+2}}$&$(y^{(r+1)})^{-\frac{2}{r+2}}D_x$&$y^{(r+1)}$\\
\hline
\vspacebeforeM
56&
${S_{r+4}}{Q_{r+3}^{-3/2}}$& ${ y^{(r+1)}}{{Q_{r+3}^{-1/2}}}D_x$& $y^{(r+1)}Q_{r+3}$\\
\hline

\end{tabular}
\end{center}

%\newpage

\subsection*{Acknowledgements}
The author is grateful to Profs.\ V.~Boyko, A.~Nikitin and R.~Popovych
for statement of the problem, helpful discussions and useful censorious remarks.
The research was supported by INTAS in the form of
PhD Fellowship (the INTAS Ref. Nr 04-83-3217) and
partially by the Grant of the President of Ukraine
for young scientists. The author thanks the anonymous
referees for helpful suggestions and comments.

\end{document}